\documentclass[preprint,aps,prd,showpacs,nofootinbib,superscriptaddress,floatfix]{revtex4-1}

\usepackage{amsmath}
\usepackage{graphicx}
\usepackage{epstopdf}
\usepackage{epsfig}
\usepackage{amssymb}
\usepackage{bm}
\newcommand{\beq}{\begin{equation}}
\newcommand{\eeq}{\end{equation}}

\usepackage{color}

\begin{document}

\begin{flushright}
\end{flushright}

\title{ Spontaneous SUSY breaking without R-symmetry in supergravity }

\author{Nobuhiro Maekawa}
\email[]{maekawa@eken.phys.nagoya-u.ac.jp }
\affiliation{Department of Physics,
Nagoya University, Nagoya 464-8602, Japan}
\affiliation{Kobayashi-Maskawa Institute for the Origin of Particles and the
Universe, Nagoya University, Nagoya 464-8602, Japan}

\author{Yuji Omura}
\email[]{yujiomur@kmi.nagoya-u.ac.jp}
\affiliation{Kobayashi-Maskawa Institute for the Origin of Particles and the
Universe, Nagoya University, Nagoya 464-8602, Japan}

\author{Yoshihiro Shigekami}
\email[]{sigekami@eken.phys.nagoya-u.ac.jp}
\affiliation{Department of Physics,
Nagoya University, Nagoya 464-8602, Japan}

\author{Manabu Yoshida}
\email[]{manabu@eken.phys.nagoya-u.ac.jp}
\affiliation{Department of Physics,
Nagoya University, Nagoya 464-8602, Japan}

\date{\today}


\begin{abstract}
\noindent
We discuss spontaneous supersymmetry (SUSY) breaking in a model with an anomalous $U(1)_A$ symmetry.
In this model, the size of the each term in the superpotential is controlled by the $U(1)_A$ charge assignment and SUSY is spontaneously broken via the Fayet-Iliopoulos of $U(1)_A$ at the meta-stable vacuum.
In the global SUSY analysis, the gaugino masses become much smaller than the sfermion masses, because an approximate R-symmetry appears at the SUSY breaking vacuum. In this paper, we show that
gaugino masses can be as large as gravitino mass, taking the supergravity effect into consideration.
This is because the R-symmetry is not imposed so that the constant term in the superpotential,
which is irrelevant to the global SUSY analysis, largely contributes to the soft SUSY breaking terms in the supergravity.
As the mediation mechanism, we introduce the contributions of the field not charged under $U(1)_A$ and the moduli  
field to cancel the anomaly of $U(1)_A$. We comment on the application of our SUSY breaking scenario to the GUT models.

\end{abstract}


\maketitle

\section{Introduction}
The conditions for spontaneous supersymmetry (SUSY) breaking have been pointed
out in the literatures \cite{Witten:1982df,Nelson:1993nf}. 
Nelson and Seiberg \cite{Nelson:1993nf} speculated that 
without R-symmetry SUSY cannot be broken spontaneously in global minimum of the scalar potential
with generic interactions, and no counter
example for this speculation has been known. On the other hand, 
since R-symmetry forbids gaugino
and higgsino masses, the R-symmetry must be broken to obtain realistic models.
However, spontaneous R-symmetry breaking results in massless R-axion which is potentially suffering from astrophysical problems. In the supergravity (SUGRA),
R-symmetry can be broken by constant term in the superpotential
without changing the arguments in global SUSY (and in many cases,
it needs to obtain Minkowski space-time \cite{Omura}), and it gives the R-axion massive \cite{R-axion}.
However, once such a R-symmetry breaking term is introduced, we have no reason
to keep R-symmetry only in SUSY breaking sector.

One solution is to break R-symmetry explicitly, although the SUSY breaking vacua
become meta-stable \cite{Omura,Kim:2008kw,meta1,Intriligator:2007py,meta2,meta3,Dienes:2008gj,Abe:2007ki}. 
In Ref. \cite{Kim:2008kw}, we examined a simple SUSY breaking model without R-symmetry, which
has following features:
\begin{enumerate}
\item Since all interactions which are allowed by the anomalous $U(1)_A$ gauge 
symmetry, which has Fayet-Iliopoulos (FI) $D$-term \cite{FI}, are introduced with $O(1)$ 
coefficients, R-symmetry is maximally broken.
\item SUSY is spontaneously broken in meta-stable vacua, at which approximate
R-symmetry appears.
\item 
Massless R-axion does not appear because of
the explicit R-symmetry breaking terms.
\end{enumerate}
Unfortunately, it seems to be difficult to apply this SUSY breaking model to
realistic scenario since the gaugino masses become much smaller than the sfermion
masses because of the approximate R-symmetry. 
The gaugino masses explicitly break the R-symmetry. 
Even if the R-symmetry is not the one of the fundamental
symmetry as in our model, the gaugino mass is vanishing because of the accidental R-symmetry at the SUSY breaking vacuum.


In this paper, we will point out that the sizable gaugino masses can be produced if 
the SUGRA effects are taken into account.
The essential point is that the constant term in the superpotential, which is irrelevant to the global SUSY analysis, contributes to the SUSY breaking dynamics. 
This term breaks not only R-symmetry explicitly, but also contributes to the SUSY breaking terms. 
As the result, generically, the gaugino masses become of order the gravitino mass, mediated by the extra fields, such as moduli fields. That is nothing but the usual results of the gravity mediation. 
However, since the vacuum structure is modified by including the
SUGRA effects and the application of the obtained model is important, we will
stress in this paper that even with approximate R-symmetry in global SUSY 
calculation, the gaugino
masses can be around the gravitino mass when the SUGRA effects are included.

In section II, we review the simple SUSY breaking model without R-symmetry.
In section III, we discuss the SUSY breaking with the SUGRA effects in the model. In section IV,
we have summary and discussion.

\section{Review of SUSY breaking model without R-symmetry}
The SUSY breaking with the FI term has been studied
in the global SUSY \cite{po,Luo:2008zr,Matos:2009xv,Dumitrescu:2010ca,Azeyanagi:2011uc,Azeyanagi:2012pc,
Vaknin:2014fxa,Kobayashi:2017fgl}.
In this section, we give a brief review of a SUSY breaking model without R-symmetry
in which the FI term of anomalous $U(1)_A$ gauge symmetry plays an 
important role in breaking SUSY spontaneously, following Ref. \cite{Kim:2008kw}.

First of all, we remind you of a simple FI model \cite{po} which has R-symmetry. 
In this model, there are two fields, $S$ and $\Theta$ whose $U(1)_A$ charges are $s\gg 1$ and 
$\theta=-1$, respectively. Note that the large charge, $s$, realizes the hierarchy between the
SUSY breaking scale and the cut-off scale (Planck scale).
If the R-charges of $S$ and $\Theta$ are 2 and 0, 
respectively, the generic superpotential is given as
\begin{equation}
W = y\Lambda^3\frac{S}{\Lambda} \left(\frac{\Theta}{\Lambda}\right)^s 
\label{eqn:51}
\end{equation}
where $y$ and $\Lambda$ are the coefficient and the cutoff of the model. 
The potential is given as
\begin{equation}
V=|F_S|^2+|F_\Theta|^2+\frac{1}{2}D_A^2,
\end{equation}
where the $F$-terms and the $D$-term are 
\begin{eqnarray}
F_S^*&=&-\frac{\partial W}{\partial S}=-y\Theta^s \\
F_\Theta^*&=&-\frac{\partial W}{\partial \Theta}=-ysS\Theta^{s-1}, \\
D_A&=&-g(\xi^2-|\Theta|^2+s|S|^2),
\end{eqnarray} 
when the K\"ahler potential $K$ is canonical. 
Here, $g$ and $\xi^2$ are the gauge coupling constant of the $U(1)_A$ gauge symmetry and
the FI parameter of the FI term, respectively. Note that we usually take $\Lambda=1$ for simplicity in this paper.
The  vacuum expectation values (VEVs) of these fields are determined by the 
minimization of the potential as
\begin{equation}
\langle S\rangle=0, \quad\langle \Theta\rangle\equiv \lambda\sim \frac{\xi}{\Lambda}.
\label{vacuaR}
\end{equation}
Note that R-symmetry is not broken at all although SUSY is broken 
spontaneously. At this vacuum, the VEVs of the $F$-terms and the $D$-term are given as
\begin{equation}
\langle F_S\rangle\sim \lambda^s, \quad \langle F_\Theta\rangle=0, \quad 
\langle D_A\rangle\sim \frac{s}{g}\lambda^{2s-2}.
\end{equation}    
$\lambda$ is expected to be ${\cal O}(0.1)$ \cite{Kim:2008kw}, so that
the SUSY breaking scale, that is given by the $F$-term and the $D$-term, becomes much smaller than the cut-off scale.

Second, we consider another model where the R-symmetry is not imposed to the above setup \cite{Kim:2008kw}. 
Then the generic superpotential is given by
\begin{equation}
W=W(S\Theta^s),
\label{superpotential}
\end{equation}
where $W(x)$ is a function of $x$ and expected to be a polynomial function as
$W(x)=\sum_{n=0}a_nx^n$. The coefficients $a_n$ are expected to be
of order one generically. 
Then SUSY vacua appear because
all of the $F$-terms and the $D$-term,
\begin{eqnarray}
F_S^*&=&-W'(S\Theta^s)\Theta^s \\
F_\Theta^*&=&-W'(S\Theta^s)sS\Theta^{s-1}, \\
D_A&=&-g(\xi^2-|\Theta|^2+s|S|^2),
\end{eqnarray} 
can be vanishing at the same time. Here, $W'(x)\equiv \frac{dW}{dx}$ is defined.
Indeed, $W'(S\Theta^s)=0$ and $D_A=0$ can be satisfied by fixing two
variables,  $\langle S\rangle$ and $\langle\Theta\rangle$, which become 
of order one generically. On the other hand, as pointed out in Ref. \cite{Kim:2008kw}, this model has
meta-stable vacua where SUSY is spontaneously broken.
The meta-stable vacua are near the vacua with the R-symmetry in Eq. (\ref{vacuaR})
as 
\begin{equation}
\langle S\rangle\sim \frac{\Theta^{s+2}|a_2|}{s^2|a_1|}\sim\frac{1}{s^2}\lambda^{s+2}, \quad\langle \Theta\rangle\equiv \lambda\sim \frac{\xi}{\Lambda}.
\label{vacuaNR}
\end{equation}
The VEVs of $F$ and $D_A$ are given as
\beq
\langle F_S\rangle\sim \lambda^s,\quad \langle F_\Theta\rangle\sim \frac{1}{s}\lambda^{2s+1}, \quad \langle gD_A\rangle\sim s\lambda^{2s-2}.
\eeq
It is obvious that the vacua have an approximate R-symmetry
when $\lambda\ll1$ and $s\gg 1$. 
Note that the VEV of $S$ is roughly proportional to the R-symmetry breaking 
parameter $a_2$. Thus, the soft SUSY breaking terms that break the R-symmetry, e.g. the gaugino masses,
become
quite small if this model is applied to the realistic models.

This is the conclusion, based on the global SUSY analysis, where the constant term in the superpotential,
$a_0$, is ignored. 
When the SUGRA effect is taken into consideration, 
we can expect that $a_0$ largely contributes to the gaugino masses. 
In the next section, we discuss this model where the R-symmetry is not imposed in the SUGRA.

\section{SUGRA effects}
In this section, we will show that SUGRA effects are not negligible especially in the models with approximate R-symmetry 
as in the previous section.
The essential point is that the constant term $a_0$ of the superpotential, 
which breaks R-symmetry, contributes to the vacua in SUGRA calculation, but not in 
global SUSY calculation. The VEV $\langle S\rangle$ in 
SUGRA calculation is proportional to $a_0$, which is much larger than $\langle S\rangle$ in global SUSY calculation in Eq. (\ref{vacuaNR}). Therefore, 
the breaking effect of $U(1)_R$ is larger at vacua
in SUGRA calculation than in global SUSY calculation.
Moreover, if there is at least one $U(1)_A$ singlet field, 
then the $F$ component of the singlet field can become sizable because the constant superpotential
contributes to the $F$ component of the singlet field. Since the singlet field can couple to 
superfield strength, the non-vanishing $F$ of the singlet can contribute to gaugino
masses.  
In addition, the $F$ component of the moduli
field, that is required to cancel the gauge anomaly in anomalous $U(1)_A$ gauge
theory, can have non-vanishing VEV because it includes the term proportional to 
the VEV of superpotential. 
These contributions can give the gaugino masses around the gravitino mass.

The superpotential is the same as in Eq. (\ref{superpotential}), although 
$a_0$ is determined by $\langle V\rangle=0$ and therefore the gravitino mass
$m_{3/2}$ is fixed by $\langle V\rangle=0$ because $a_0=m_{3/2}M_{Pl}^2$.
Here, 
$M_{Pl}$ is 
the reduced Planck scale.
We treat the cutoff scale $\Lambda$ and the Planck scale differently, as in 
Horava-Witten theory \cite{HW} or in natural GUT \cite{naturalGUT}.
Then the scalar potential is written as
\begin{eqnarray}
V&=&V_F +V_D, \nonumber \\
V_F &=& e^ {K/M_{Pl}^2} \left ( |D_S W|^2 +  |D_\Theta W|^2 -3 \frac{|W|^2}{M_{Pl}^2} \right ),  \\
V_D &=& \frac{g^2}{2} \left ( \xi^2 +s |S|^2-|\Theta|^2 \right )^2,
\end{eqnarray}
where the following functions are defined:
\begin{eqnarray}
K&=& |S|^2 +|\Theta|^2, \\
D_S W &=& W'(S\Theta^s)\Theta^s+ \frac{S^*}{M^2_{Pl}} W ,  \\
D_\Theta W &=&W'(S\Theta^s)sS\Theta^{s-1}+ \frac{\Theta^*}{M^2_{Pl}} W.
\end{eqnarray}
The stationary conditions for the potential give the VEVs of $S$ and $\Theta$ as
\beq
\langle \Theta\rangle\sim \lambda, \quad \langle S\rangle\sim \frac{\lambda^2}{s}\frac{\Lambda}{M_{Pl}},
\eeq
and the vanishing cosmological constant $\langle V\rangle=0$ fixes 
$\langle W\rangle\sim \lambda^sM_{Pl}$, which determines the gravitino mass $m_{3/2}$ as
 $m_{3/2}=\langle W\rangle/M_{Pl}^2\sim \lambda^s\frac{\Lambda}{M_{Pl}}$.
The VEVs of $D_S W$, $D_\Theta W$,  and $D_A$ are given as
\beq
\langle D_SW\rangle\sim \lambda^s,\quad \langle D_\Theta W\rangle\sim \lambda^{s+1}\frac{\Lambda}{M_{Pl}}, \quad \langle gD_A\rangle\sim s\lambda^{2s-2}.
\eeq
Note that $\langle S\rangle$ is not so small at all especially when $\Lambda\sim M_{Pl}$, and therefore
the VEVs break R-symmetry completely. This may induce gaugino masses if this
mechanism is embedded in realistic model. When $S=S_re^{i\frac{\phi_S}{\sqrt{2}\langle S\rangle}}$, the masses of $S_r$, $\phi_S$ and $\Theta$ are given as $\frac{sM_{Pl}}{\lambda\Lambda}m_{3/2}$, $\frac{sM_{Pl}}{\lambda \Lambda}m_{3/2}$, and $\Lambda$, respectively.

Moreover, the $F$ component of a field $Z$ which is neutral under
$U(1)_A$ can have non-vanishing because of the contribution 
from the constant superpotential. This also gives sizable gaugino masses which can be
around the gravitino mass. We show this in an explicit model in which the neutral field $Z$ is
added  to  the above model. 
Then the scalar potential is written as
\begin{eqnarray}
V&=&V_F +V_D, \nonumber \\
V_F &=& e^ {K/M_{Pl}^2} \left ( |D_S W|^2 +  |D_\Theta W|^2+|D_ZW|^2 -3 \frac{|W|^2}{M_{Pl}^2} \right ),  \\
V_D &=& \frac{g^2}{2} \left ( \xi^2 +s |S|^2-|\Theta|^2 \right )^2,
\end{eqnarray}
where the following functions are defined:
\begin{eqnarray}
K&=&|S|^2 +|\Theta|^2+|Z|^2, \\
D_S W &=& W'(S\Theta^s)\Theta^s+ \frac{S^*}{M^2_{Pl}} W ,  \\
D_\Theta W &=&W'(S\Theta^s)sS\Theta^{s-1}+ \frac{\Theta^*}{M^2_{Pl}} W, \\
D_ZW&=&\dot W+\frac{Z^*}{M_{Pl}^2}W,
\end{eqnarray}
where the superpotential is given as
\beq
W=\sum_{n=0}a_n(Z)(S\Theta^s)^n=W(S\Theta^s).
\eeq
Here, $W'=\frac{dW(x)}{dx}$ and $\dot W=\frac{\partial W}{\partial Z}=\sum_{n=0}\frac{da_n}{dZ}(S\Theta^s)^n$. 
The VEVs are essentially the same as the previous results except the VEV
of $Z$. 
The stationary condition $\partial V/\partial Z=0$ determines the VEV of
$D_ZW$ as
\beq
\langle D_ZW\rangle\sim \frac{\langle\dot W'\rangle}{\langle W'\rangle\langle\ddot W\rangle}m_{3/2}^2M_{Pl}^2\sim \frac{\dot a_1(\langle Z\rangle)}{a_1(\langle Z\rangle)\langle\ddot a_0\rangle}m_{3/2}^2M_{Pl}^2.
\eeq
Since $\dot a_1\sim a_1$ is expected, $\langle D_ZW\rangle$ becomes
\beq
\langle D_ZW\rangle\sim\left\{\begin{array}{l} m_{3/2} \quad ({\rm when} \ \langle \ddot a_0\rangle\sim m_{3/2}M_{Pl}^2)\cr
m_{3/2}^2M_{Pl}^2\quad ({\rm when}\ \langle \ddot a_0\rangle\sim 1)
\end{array}\right.
\label{FZ}
\eeq
Note that the VEV $\ddot W$ is dependent on the mechanism to realize
$\langle V\rangle=0$. For example, if $a_0(Z)=m_{3/2}M_{Pl}^2\hat a_0(Z)$, where
$\hat a_0(Z)$ is a polynomial function with $O(1)$ coefficients, then the upper result
in Eq. (\ref{FZ}) is realized, and the mass of $Z$ becomes $\frac{M_{Pl}^2}{\Lambda^2}m_{3/2}$. If $a_0(Z)$ is a polynomial function with
$O(1)$ coefficient whose VEV is $\langle a_0\rangle=m_{3/2}M_{Pl}^2$, 
the lower result is realized and the mass of $Z$ becomes $\Lambda$. 
 It is important that the VEV of $D_ZW$ can
be $O(m_{3/2})$ and therefore, gaugino masses can be $O(m_{3/2})$ because
the neutral field $Z$ can couple with the kinetic functions of vector multiplets.

There is another contribution to gaugino masses from the $F$-term of
the moduli fields $T$, that can be $O(m_{3/2})$. Since $U(1)_A$
gauge symmetry is given by
\begin{eqnarray}
V_A&\rightarrow& V_A+\frac{i}{2}(\tilde \Lambda-\tilde \Lambda^\dagger), \\
T&\rightarrow &T+\frac{i}{2}\delta_{\rm GS}\tilde\Lambda,
\end{eqnarray}
where $V_A$, $\tilde \Lambda$, and $\delta_{\rm GS}$ are vector multiplet
of $U(1)_A$, a gauge parameter chiral superfield, and dimensionless parameter which has relations\footnote{
The relations can be satisfied by choosing normalization factor of
$U(1)_A$ gauge symmetry and/or $k_A$, although we do not fix these
explicitly in this paper.
}
\beq
2\pi^2\delta_{\rm GS}=\frac{1}{3k_A}{\rm tr}Q_A^3=\frac{1}{24}{\rm tr} Q_A>0.
\eeq
The anomaly of $U(1)_A$
 can be cancelled \cite{GS} via
\beq
{\cal L}_{\rm gauge}=\frac{1}{4}\int d\theta^2k_ATW_A^\alpha W_{A\alpha}+h.c.,
\eeq
where $W_A^\alpha$ and $k_A$ are the super field strength of $V_A$ and
Kac-Moody level of $U(1)_A$, respectively. The $U(1)_A$ invariant K\"ahler
potential is given as 
\beq
{\cal K}=S^\dagger e^{-2gsV_A}S +\Theta^\dagger e^{2gV_A}\Theta+f(T+T^\dagger-\delta_{\rm GS}V_A).
\eeq
The FI term can be given as
\beq
\int d^4\theta f(T+T^\dagger-\delta_{\rm GS}V_A)=\left(-\frac{\delta_{\rm GS}f'}{2}\right)D_A\cdots\equiv \xi^2D_A+\cdots.
\eeq
Note that $\langle f'\rangle$ must be negative to obtain positive $\xi^2$.

The scalar potential
is given as
\begin{eqnarray}
V&=&V_F +V_D, \nonumber \\
V_F &=& e^ {K/M_{Pl}^2} \left ( |D_S W|^2 +  |D_\Theta W|^2+f''^{-1}|D_TW|^2 -3 \frac{|W|^2}{M_{Pl}^2} \right ),  \\
V_D &=& \frac{g^2}{2} \left ( -\frac{\delta_{\rm GS}f'}{2} +s |S|^2-|\Theta|^2 \right )^2,
\end{eqnarray}
where the following functions are defined:
\begin{eqnarray}
D_S W &=& W'(S\Theta^s)\Theta^s+ \frac{S^*}{M^2_{Pl}} W ,  \\
D_\Theta W &=&W'(S\Theta^s)sS\Theta^{s-1}+ \frac{\Theta^*}{M^2_{Pl}} W, \\
D_TW&=&\frac{f'}{M_{Pl}^2}W,
\end{eqnarray}
where the superpotential is given as in Eq. (\ref{superpotential}).
We expand the function $f$ around $2T_0$ as $f(T+T^\dagger)=\sum_n\frac{b_n}{n!}(T+T^\dagger-2T_0)^n$. The stationary 
condition $\partial V/\partial T=0$ gives
\beq
m_{3/2}^2\left(2f'-\frac{(f')^2f'''}{(f'')^2}\right)+gD_A\left(\frac{\delta_{\rm GS}}{2}f''\right)=0.
\eeq
The second derivative of the scalar potential becomes
\beq
\frac{\partial^2V}{\partial T^2}\sim \frac{\delta_{\rm GS}^2}{4}(f'')^2>0.
\eeq
Therefore, if 
\beq
m_{3/2}^2\left(2b_1-\frac{b_1^2b_3}{b_2^2}\right)+gD_A\left(\frac{\delta_{\rm GS}}{2}b_2\right)=0
\eeq
is satisfied, 
the VEV ${\rm Re}\langle T\rangle=T_0$ is (meta-) stable. 
Note that the moduli can easily be stabilized because of the $D$-term.
The scalar masses can be calculated as
\beq
m_T\sim\frac{\delta_{\rm GS}}{2}\Lambda, \quad m_\Theta\sim \Lambda,
\quad m_{S_r}\sim\frac{sM_{Pl}}{\lambda\Lambda}m_{3/2}, \quad
m_{\phi_S}\sim \frac{sM_{Pl}}{\lambda\Lambda}m_{3/2},  
\eeq
except massless axion.
  Actually, this scalar potential has an global $U(1)$
symmetry in addition to $U(1)_A$ gauge symmetry, which transforms only
 $S$ and $\Theta$ as $U(1)_A$ but not $T$. Because of this additional
 $U(1)$ symmetry, a Nambu-Goldstone boson appears. 
If non-perturbative interactions are allowed in the superpotential or in the  K\"ahler
potential like $\Theta e^{2T/\delta_{\rm GS}}$ or $Se^{-2sT/\delta_{\rm GS}}$
which break the additional global $U(1)$ symmetry, the axion becomes
massive. Otherwise, this axion works as QCD axion, which may solve the
strong CP problem \cite{PQ}. The effective Peccei-Quinn scale
becomes
$
F_{PQ}\sim \frac{\Lambda}{8\pi^2},
$
which is around $10^{14}$ GeV
 if $\Lambda\sim \Lambda_G\sim 2\times 10^{16}$
GeV.
The $F$-term of $T$ becomes
\beq
F_T=(f'')^{-1}D_TW=\frac{f'}{f''}\frac{W}{M_{Pl}^2},
\eeq
which gives gaugino masses as $k_A\frac{b_1}{b_2}m_{3/2}$.

In conclusion, even if the vacua determined by global SUSY calculation have approximate R-symmetry, SUGRA effects can change them to the vacua without
R-symmetry. As the result, gaugino masses can be around gravitino mass in this model.

\section{Summary and discussion}
It is one of the important issues how a realistic SUSY breaking vacuum can be realized, in supersymmetric models.
The R-symmetry seems to play an important role in the SUSY breaking, but
it causes the massless Goldstone boson and prevents generating the non-vanishing gaugino masses.
As pointed out in Ref. \cite{Intriligator:2007py}, one realistic SUSY breaking vacuum could be realized if
explicit R-symmetry breaking terms are enough small for the life time of the vacuum to be longer than the age of our universe. This scenario, however, requires the explanation of the origin of the tiny R-symmetry breaking terms.
 In addition, the gaugino actually
needs large R-symmetry breaking effects to gain large mass. \footnote{Even if the R-symmetry is spontaneously broken,
the gaugino masses are often vanishing in the gauge mediation scenario \cite{Komargodski:2009jf}.} 
Even taking the SUGRA effect into consideration, this situation does not change \cite{Omura}.
In the SUGRA, the large constant term in the superpotential, that breaks the R-symmetry,
 is necessary for the vanishing 
cosmological constant in many cases \cite{Omura}, so that the situation may become worse
 compared to the global SUSY case. Thus, we need to find the symmetry or the dynamics that can replace the role of the R-symmetry, in order to lead a realistic SUSY breaking vacuum.
We have discussed spontaneous SUSY breaking via the FI term in a model 
which has anomalous $U(1)_A$ symmetry.
The R-symmetry is not imposed, but an approximate R-symmetry appears
at the meta-stable SUSY breaking vacua in the global SUSY analysis. 
Then, the gaugino masses become much smaller than
the sfermion masses in the global SUSY as shown in Ref. \cite{Kim:2008kw}.
In this paper, we have pointed out that if the SUGRA effects are taken into account, the R-symmetry
is largely broken by the constant term in the superpotential at the meta-stable SUSY breaking vacua, 
and as the result, the gaugino masses can be of order the gravitino mass. 

In our calculation, we have adopted the cutoff scale which can be different
from the Planck scale as in natural GUT or in Horava-Witten theory.
The application of this mechanism to the natural GUT
is interesting. In the natural GUT, the doublet-triplet splitting problem can be solved under a reasonable assumption in which
all interactions including higher dimensional interactions are introduced 
with $O(1)$ coefficients. An important point is that the natural GUT has
the cutoff scale which is the usual GUT scale smaller than
the Planck scale.
As the result, sfermion masses become around 100 times larger than
the gaugino masses. Namely high scale SUSY (or split SUSY) is realized
in the model. The details will be discussed in a separate paper.

In the explicit model we discussed, we adopted an anomalous $U(1)_A$ gauge
symmetry with FI term.
 The anomaly can be cancelled by Green-Schwarz
mechanism \cite{GS} in which moduli plays an important role. We have shown explicitly that all scalar fields
become massive except a Nambu-Goldstone boson which can solve the strong CP problem.
Especially when the cutoff scale is lower than the Planck scale, these massive
modes become much heavier than the gravitino mass. 


To obtain the gaugino masses around 1 TeV which are the same order of
the gravitino mass, the gravitino problem \cite{Weinberg:1982zq,Khlopov:1984pf,Ellis:1984eq,Reno:1987qw,Kawasaki:1994af,
Kohri:2001jx,Kawasaki:2004qu,Kawasaki:2008qe} becomes serious. One possible
way to avoid the problem is to adopt low reheating temperature of inflation.
We will not discuss this problem further in this paper. 

\section{Acknowledgement}
This work is supported in part by the Grant-in-Aid for Scientific Research 
 No.~15K05048(N.M.), No. 17H05404 (Y.O.), and No.~16J08299 (Y.S.)  from the Ministry of Education, Culture, Sports,
 Science and Technology in Japan.


\begin{thebibliography}{99}

\bibitem{Witten:1982df} 
  E.~Witten,
  Nucl.\ Phys.\ B {\bf 202}, 253 (1982).






\bibitem{Nelson:1993nf}
  A.~E.~Nelson and N.~Seiberg,
  Nucl.\ Phys.\  B {\bf 416}, 46 (1994).
  [hep-ph/9309299].        


 
 \bibitem{Omura}
 H.~Abe, T.~Kobayashi and Y.~Omura,
  JHEP {\bf 0711} (2007) 044
  [arXiv:0708.3148 [hep-th]].
 
\bibitem{R-axion} 
  J.~Bagger, E.~Poppitz and L.~Randall,
  Nucl.\ Phys.\ B {\bf 426}, 3 (1994)
  [hep-ph/9405345].
 
\bibitem{Kim:2008kw} 
  S.-G.~Kim, N.~Maekawa, H.~Nishino and K.~Sakurai,
  Phys.\ Rev.\ D {\bf 79}, 055009 (2009)
  [arXiv:0810.4439 [hep-ph]].

\bibitem{meta1}
       M. Dine, A.E. Nelson, Y. Nir and Y. Shirman, Phys. Rev. {\bf D 53},
        2658, (1996)  [hep-ph/9507378]; M.A. Luty and J. Terning, Phys. Rev. {\bf D 62}, 
       075006 (2000) [hep-ph/9812290]; N. Maekawa, hep-ph/0004260; T. Banks, hep-ph/0007146.

\bibitem{meta2} 
  K.~A.~Intriligator, N.~Seiberg and D.~Shih,
  JHEP {\bf 0604}, 021 (2006)
  [hep-th/0602239].
        
\bibitem{Intriligator:2007py} 
  K.~A.~Intriligator, N.~Seiberg and D.~Shih,
  JHEP {\bf 0707}, 017 (2007)
  [hep-th/0703281].

\bibitem{meta3}
  A.~Amariti, L.~Girardello and A.~Mariotti,
  JHEP {\bf 0612} (2006) 058 [hep-th/0608063]; 
S.~A.~Abel and V.~V.~Khoze,
  arXiv:hep-ph/0701069; S.~Forste,
  Phys.\ Lett.\  B {\bf 642} (2006) 142 [hep-th/0608036]; 
M.~Gomez-Reino and C.~A.~Scrucca,
  JHEP {\bf 0708} (2007) 091 [arXiv:0706.2785[hep-th]]; 
R.~Essig, K.~Sinha and G.~Torroba,
  JHEP {\bf 0709} (2007) 032 [arXiv:0707.0007[hep-th]]; S.~Abel, C.~Durnford, J.~Jaeckel and V.~V.~Khoze,
  Phys.\ Lett.\  B {\bf 661} (2008) 201 [arXiv:0707.2958[hep-ph]]; 
A.~Giveon and D.~Kutasov,
  Nucl.\ Phys.\  B {\bf 796}, 25 (2008) [arXiv:0710.0894[hep-th]]; 
A.~Giveon, A.~Katz and Z.~Komargodski,
  JHEP {\bf 0806}, (2008), 003 [hep-th/08041805]..





\bibitem{Dienes:2008gj} 
  K.~R.~Dienes and B.~Thomas,
  Phys.\ Rev.\ D {\bf 78}, 106011 (2008)
  [arXiv:0806.3364 [hep-th]].

\bibitem{Abe:2007ki} 
  H.~Abe, T.~Kobayashi and Y.~Omura,
  Phys.\ Rev.\ D {\bf 77}, 065001 (2008)
  [arXiv:0712.2519 [hep-ph]].
\bibitem{FI}
       P.Fayet and J.Iliopoulos, Phys, Lett. {\bf B51}, 461 (1974);  
        P.Fayet, Nucl,Phys. {\bf B90}, 104 (1975).
\bibitem{po}
        G. Dvali and A. Pomarol, Phys. Rev. Lett. {\bf 77}, 3728, (1996) [hep-ph/9607383]; 
P.Binetruy and E.Dudas, Phys.Lett. {\bf B389}, 503, (1996) [hep-th/9607172].
\bibitem{Luo:2008zr} 
  M.~Luo and S.~Zheng,
  JHEP {\bf 0901}, 004 (2009)
  [arXiv:0812.4600 [hep-ph]].

\bibitem{Matos:2009xv} 
  L.~F.~Matos,
  arXiv:0910.0451 [hep-ph].


\bibitem{Dumitrescu:2010ca} 
  T.~T.~Dumitrescu, Z.~Komargodski and M.~Sudano,
  JHEP {\bf 1011}, 052 (2010)
  [arXiv:1007.5352 [hep-th]].

\bibitem{Azeyanagi:2011uc} 
  T.~Azeyanagi, T.~Kobayashi, A.~Ogasahara and K.~Yoshioka,
JHEP {\bf 1109}, 112 (2011)
[arXiv:1106.2956 [hep-ph]].
\bibitem{Azeyanagi:2012pc} 
  T.~Azeyanagi, T.~Kobayashi, A.~Ogasahara and K.~Yoshioka,
  Phys.\ Rev.\ D {\bf 86}, 095026 (2012)
  [arXiv:1208.0796 [hep-ph]].

\bibitem{Vaknin:2014fxa} 
  T.~Vaknin,
  JHEP {\bf 1409}, 004 (2014)
  [arXiv:1402.5851 [hep-th]].

\bibitem{Kobayashi:2017fgl} 
  T.~Kobayashi, Y.~Omura, O.~Seto and K.~Ueda,
  JHEP {\bf 1711}, 073 (2017)
  [arXiv:1705.00809 [hep-ph]].



\bibitem{HW}
  P.~Horava and E.~Witten,
  Nucl.\ Phys.\ B {\bf 460}, 506 (1996)
  [hep-th/9510209].
\bibitem{naturalGUT}
  N.~Maekawa,
  Prog.\ Theor.\ Phys.\  {\bf 106}, 401 (2001)
  [hep-ph/0104200];
        M. Bando and N. Maekawa, Prog. Theor. Phys. {\bf 106}, 1255 (2001).
  Prog.\ Theor.\ Phys.\  {\bf 107}, 597 (2002)
  [hep-ph/0111205];
  N.~Maekawa and T.~Yamashita,
  Phys.\ Rev.\ Lett.\  {\bf 90}, 121801 (2003)
  [hep-ph/0209217];
  N.~Maekawa and T.~Yamashita,
  Prog.\ Theor.\ Phys.\  {\bf 107}, 1201 (2002)
  [hep-ph/0202050].





\bibitem{GS} 
  M.~B.~Green and J.~H.~Schwarz,
  Phys.\ Lett.\  {\bf 149B}, 117 (1984).

\bibitem{PQ}
  R.~D.~Peccei and H.~R.~Quinn,
  Phys.\ Rev.\ Lett.\  {\bf 38}, 1440 (1977).




\bibitem{Komargodski:2009jf} 
  Z.~Komargodski and D.~Shih,
JHEP {\bf 0904}, 093 (2009)
[arXiv:0902.0030 [hep-th]].






\bibitem{Weinberg:1982zq}
  S.~Weinberg,
  Phys.\ Rev.\ Lett.\  {\bf 48} (1982) 1303.


\bibitem{Khlopov:1984pf}
  M.~Y.~Khlopov and A.~D.~Linde,
  Phys.\ Lett.\ B {\bf 138} (1984) 265.


\bibitem{Ellis:1984eq}
  J.~R.~Ellis, J.~E.~Kim and D.~V.~Nanopoulos,
  Phys.\ Lett.\ B {\bf 145} (1984) 181.


\bibitem{Reno:1987qw}
  M.~H.~Reno and D.~Seckel,
  Phys.\ Rev.\ D {\bf 37} (1988) 3441.


\bibitem{Kawasaki:1994af}
  M.~Kawasaki and T.~Moroi,
  Prog.\ Theor.\ Phys.\  {\bf 93} (1995) 879
  [hep-ph/9403364, hep-ph/9403061].


\bibitem{Kohri:2001jx}
  K.~Kohri,
  Phys.\ Rev.\ D {\bf 64} (2001) 043515
  [astro-ph/0103411].


\bibitem{Kawasaki:2004qu}
  M.~Kawasaki, K.~Kohri and T.~Moroi,
  Phys.\ Rev.\ D {\bf 71} (2005) 083502
  [astro-ph/0408426].


\bibitem{Kawasaki:2008qe}
  M.~Kawasaki, K.~Kohri, T.~Moroi and A.~Yotsuyanagi,
  Phys.\ Rev.\ D {\bf 78} (2008) 065011
  [arXiv:0804.3745 [hep-ph]].



\end{thebibliography}
\end{document}